\documentclass[reprint, twocolumn, aps, pra, floatfix,superscriptaddress,longbibliography]{revtex4-2}

\usepackage{graphicx} 
\usepackage{float} 

\usepackage{xcolor} 
\usepackage{amsmath} 
\usepackage{amssymb}

\usepackage{hyperref}

\begin{document}
 
\title{Dark-fringe interferometer  with dynamic phase control for M\"ossbauer science}

\author{Miriam Gerharz}
\affiliation{Max-Planck-Institut f\"ur Kernphysik, Saupfercheckweg 1, 69117 Heidelberg, Germany}

\author{Dominik Lentrodt}
\affiliation{Max-Planck-Institut f\"ur Kernphysik, Saupfercheckweg 1, 69117 Heidelberg, Germany}

\author{Lars Bocklage}
\affiliation{Deutsches Elektronen-Synchrotron DESY, Notkestraße 85, 22607 Hamburg, Germany}
\affiliation{The Hamburg Centre for Ultrafast Imaging CUI, Luruper Chaussee 149, 22761 Hamburg, Germany}

\author{Kai Schulze}
\affiliation{Helmholtz-Institut Jena, Fr\"obelstieg 3, 07743 Jena, Germany}
\affiliation{Institut f\"ur Optik und Quantenelektronik, Friedrich-Schiller-Universit\"at Jena, Max-Wien-Platz 1, 07743 Jena, Germany}

\author{Christian Ott}
\affiliation{Max-Planck-Institut f\"ur Kernphysik, Saupfercheckweg 1, 69117 Heidelberg, Germany}

\author{Ren\'e Steinbr\"ugge}
\affiliation{Deutsches Elektronen-Synchrotron DESY, Notkestraße 85, 22607 Hamburg, Germany}

\author{Olaf Leupold}
\affiliation{Deutsches Elektronen-Synchrotron DESY, Notkestraße 85, 22607 Hamburg, Germany}

\author{Ilya Sergeev}
\affiliation{Deutsches Elektronen-Synchrotron DESY, Notkestraße 85, 22607 Hamburg, Germany}

\author{Gerhard G. Paulus}
\affiliation{Helmholtz-Institut Jena, Fr\"obelstieg 3, 07743 Jena, Germany}
\affiliation{Institut f\"ur Optik und Quantenelektronik, Friedrich-Schiller-Universit\"at Jena, Max-Wien-Platz 1, 07743 Jena, Germany}

\author{Christoph H. Keitel}
\affiliation{Max-Planck-Institut f\"ur Kernphysik, Saupfercheckweg 1, 69117 Heidelberg, Germany}

\author{Ralf R\"ohlsberger}
\affiliation{Deutsches Elektronen-Synchrotron DESY, Notkestraße 85, 22607 Hamburg, Germany}
\affiliation{The Hamburg Centre for Ultrafast Imaging CUI, Luruper Chaussee 149, 22761 Hamburg, Germany}
\affiliation{Helmholtz-Institut Jena, Fr\"obelstieg 3, 07743 Jena, Germany}
\affiliation{Institut f\"ur Optik und Quantenelektronik, Friedrich-Schiller-Universit\"at Jena, Max-Wien-Platz 1, 07743 Jena, Germany}
\affiliation{Helmholtz Centre for Heavy Ion Research (GSI), Planckstr. 1, 64291 Darmstadt, Germany}

\author{Thomas Pfeifer}
\affiliation{Max-Planck-Institut f\"ur Kernphysik, Saupfercheckweg 1, 69117 Heidelberg, Germany}

\author{J\"org Evers}
\email{joerg.evers@mpi-hd.mpg.de}
\affiliation{Max-Planck-Institut f\"ur Kernphysik, Saupfercheckweg 1, 69117 Heidelberg, Germany}

\newcommand{\circpol}[0]{\hat\sigma} 
\newcommand{\perppol}[0]{\hat\sigma} 
\newcommand{\parapol}[0]{\hat\pi} 

\newcommand{\Rlin}{R_{\textrm{L}}}
\newcommand{\Rcirc}{R_{\textrm{C}}}
\newcommand{\Tlin}{T_{\textrm{L}}}
\newcommand{\Tcirc}{T_{\textrm{C}}}
\newcommand{\Rlc}{R_\textrm{L/C}}
\newcommand{\Tlc}{T_\textrm{L/C}}
\newcommand{\je}[1]{{\color{red}{#1}}}


\begin{abstract} 
    Interference is a powerful tool for measuring and control. In M\"ossbauer science, interference effects are essential to most applications, due to the coherent scattering nature. However, M\"ossbauer interferometry remains challenging, due to stability requirements imposed by the short x-ray wavelength. Here, we put forward a ``dark fringe'' interferometer with vanishing transmission in the empty state, thereby facilitating sensitive measurements. The  relative interferometer phase can dynamically be tuned by displacing a M\"ossbauer target. We experimentally demonstrate the tuning capabilities of this interferometer by controlling the transmitted x-ray intensity on  nanosecond time scales. Then, we demonstrate sensitive measurements by observing  the propagation of impulsively launched sound waves in the target over $\sim 10\,\mu s$. The interferometer concept opens avenues towards polarization-sensitive phase measurements, the generation of coherent multi-pulse sequences for controlling nuclear dynamics, and the implementation of feedback loops to adaptively optimize the interferometer, thereby fueling the further development of nuclear quantum optics.
\end{abstract}

\maketitle

\section{Introduction}

The phenomenon of interference is ubiquitous in physics. For instance, the paradigmatic example of double-slit interference allows one to explore key concepts of quantum mechanics~\cite{Feynman}. In general, interferometers are among the most versatile and precise measurement devices. However, interferometric control and measurements become more challenging at energies of hard x-rays, due to the small x-ray wavelength. Nevertheless, interference effects also play an important role in nuclear resonance scattering, which typically operates at photon energies of order 10~keV. Already the interference between scattering involving different hyperfine transitions  gives rise to the characteristic quantum beats~\cite{ralf}. The situation becomes richer if more than one nuclear target is considered, since then also interferences between pathways involving interactions with several targets become relevant (see, e.g.,~\cite{PhysRevA.61.013803,PhysRevA.63.043810,SciencePaper_PiezoPhase,heeg2021coherent}). These interferences open up a number of important applications, such as controlling the scattered $\gamma$-rays and the nuclear dynamics~\cite{PhysRevA.61.013803,PhysRevA.63.043810,Shvyd_ko_1989,Helistoe1991,PhysRevLett.77.3232,PhysRevLett.82.3593,Schindelmann2002,PhysRevLett.103.017401,PhysRevLett.109.197403,PhysRevLett.111.073601,Vagizov2014,PhysRevLett.114.207401,SciencePaper_PiezoPhase,Liao2017,PhysRevLett.123.250504,heeg2021coherent,bocklage2021.20097,PhysRevApplied.18.L051001,PhysRevLett.133.193401,doi:10.1126/sciadv.adn9825} or in precise measurements~\cite{PhysRevB.63.094105,refId0,PhysRevB.72.081402,PhysRevLett.123.153902,heeg2021coherent,bocklage2021.20097,Yuan2025,lohse2025interferometricmeasurementnuclearresonant}. 
One example is the manipulation of the intensity of the transmitted light, e.g., motivated by the desire to remove the huge off-resonant unscattered background from the detection signal at accelerator-based x-ray sources~\cite{PhysRevLett.70.359,doi:10.1063/1.114764,PhysRevLett.111.073601,Chechin,PhysRevLett.54.835,Smirnov2000,Potapkin:vv5038,R_hlsberger_1992,Smirnov1984,Shvyd_ko_1989,PhysRevLett.84.1007,Toellner:ie5046,argonne-mems}.
From a broader perspective, the M\"ossbauer interference schemes  are also at the heart of the development of x-ray and nuclear quantum optics~\cite{Adams2013,Rohlsberger2014,kuznetsova2017quantum,adams_scientific_2019old,wong2021prospects,Adams2003}.

The favorable properties and applications call for a further development of M\"ossbauer interferometry. However, traditional x-ray interferometers based on crystal optics are challenging to implement, since they require an alignment and stability of the setup on the Angstrom level~\cite{Bonse1965,Shvydko2003,Bowen1996}. In contrast, in the multi-target M\"ossbauer settings the interfering pathways are not spatially separated. This has the advantage that the measurements are only susceptible to relative phase changes between the targets over the duration between the excitation and the subsequent scattering.  For the archetype ${}^{57}$Fe with a lifetime of 141~ns, this measurement cycle is fast enough to outpace many sources of mechanical noise. As a result, stabilities well below the wavelength scale have been demonstrated~\cite{heeg2021coherent}. However, in a conventional interferometer the probed sample is placed in one of the interfering pathways, while a variable phase shift is applied to the other pathway. Neither the selective coupling to one of the paths, nor the phase variation is  straightforwardly possible in the inline configuration. Furthermore, the radiative coupling between multiple  M\"ossbauer targets~\cite{PhysRevA.61.013803,PhysRevA.63.043810} involving x-ray scattering on more than one target spoils the analogy to traditional interferometry setups, and impedes the direct phase measurements and control~\cite{PhysRevB.72.081402,PhysRevB.73.184126,heeg2021coherent,PhysRevLett.123.153902}. 

Here, we theoretically introduce and experimentally demonstrate a dynamically-controllable  x-ray interferometer for M\"ossbauer science which overcomes these challenges. We engineer the interference in such a way that only two co-propagating pathways contribute to the detection signal. In this setting, the radiative couplings cancel, the two pathways can individually be addressed via their  different polarizations, and the transmitted intensity in the empty interferometer ideally vanishes. Despite not being used in Mössbauer interferometers so far, due to its high sensitivity, this ``dark-fringe'' mode is  well-established in other fields, e.g. in astrophysics \cite{Hinz1998} or gravitational-wave detection \cite{Bond2016}. 
The interferometer is composed of two $\alpha$-iron targets enriched in ${}^{57}$Fe, placed inside a high-purity polarimetry setup. We show that starting from the dark-fringe setting, a relative displacement of the two targets allows one to continuously tune the relative phase between the two interfering pathways, thereby completing the analogy to a traditional interferometry setup. 
To experimentally demonstrate the capabilities of the interferometer,  we use an event-based detection system to measure the effect of individual sudden target displacements on the interferometer transmission up to about 7.5~$\mu$s, i.e., about 50 lifetimes of the bare nuclei. As the first application, we then demonstrate a control of the transmitted x-ray intensity on the nanosecond time scale using suitable displacements of one of the targets. Second, we demonstrate a measurement of relative displacements of the two targets on x-ray wavelength- and nanosecond-time scales, by observing the propagation and reflection of mechanical waves induced in the target holder by a sudden target displacements over durations of more than 10$\mu s$. 

\begin{figure}[t]
    \centering
    \includegraphics[width = \columnwidth]{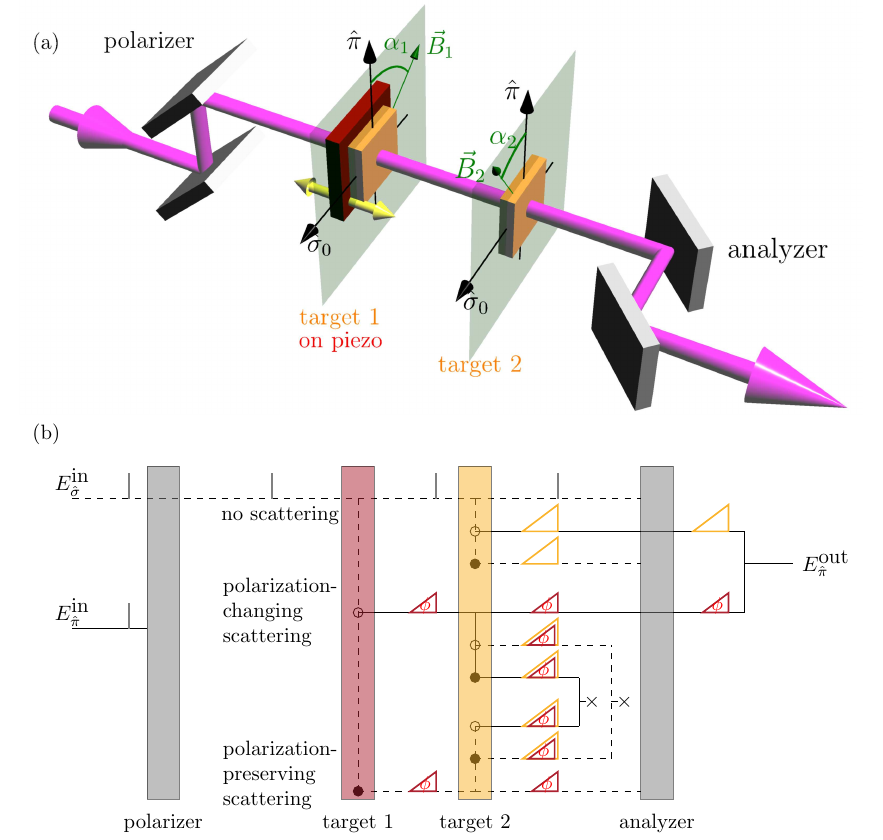}
    \caption{(a) Schematic setup. X-rays linearly polarized in ${\perppol}$-direction enter the setup. With a polarimetry setup, the two perpendicular polarization states $E_{\perppol}$ and $E_{\parapol}$ can be selected. Inside the polarimeter, two resonant targets are placed with their magnetization set to {$\alpha_1=\pi/4$ and $\alpha_2=-\pi/4$}, which forms the dark fringe setting, in which no light can pass the setup. The first target is mechanically moved via a piezoelectric transducer and thereby imprints a time-dependent phase onto the x-rays. This allows to dynamically control the interferometer and enables a dynamical temporal gating of the  outgoing x-ray intensity. 
    (b) Schematic representation of the different interfering paths. First, the polarizer blocks the $\hat{\pi}$-component (solid line) and lets only pass the $\hat{\sigma}$-component (dashed line) of the initial short x-ray pulse (gray vertical line) of the incoming electric field. In the first target, the light can either pass the target without scattering (top path) or is scattered into the perpendicular component (empty circle) or into the same polarization component (filled circle). Both scattering components are indicated by dark red triangles since their duration is of order of the nuclear lifetime and therefore orders of magnitude longer than the initial x-ray pulse. They have a dynamically-controllable phase $\phi$ relative to the initial x-ray pulse. Subsequently, similar scattering  happens in the second target (orange triangle). The two double-scattering paths interfere destructively for both polarization components (crossed arrow). Finally, the analyzer blocks the $\hat{\sigma}$ components, such that the outgoing electric field behind the interferometer is given by the sum of two single-scattering paths, whose relative phase can be controlled.
    }
    \label{fig:setup}
\end{figure}

\section{Results}
\subsection{Dark-fringe interferometer}
We start by explaining the setup and the operation principle of the dark-fringe interferometer, see Fig.~\ref{fig:setup}(a). In our interferometer, the interfering pathways differ in their polarization evolution while they pass through the setup. The corresponding polarization analysis is facilitated by a polarimetry setup~\cite{MARX2011915,Marx-Glowna2021}, which polarizes the incoming x-rays in the $\hat \sigma$ direction. The subsequent analyzer is aligned in crossed setting, such that the empty polarimeter ideally does not transmit any x-rays. 
To harness the polarization degree of freedom, we place two targets into the polarimeter, which feature scattering between the $\hat \sigma$ and the $\hat \pi$ polarization directions.
The targets contain  the isotope ${}^{57}$Fe with resonance frequency $\omega_0$ and a magnetic dipole (M1) M\"ossbauer transition from the ground state. This transition has a hyperfine splitting into six lines of which two are linearly polarized along the magnetization and four are circularly polarized in the plane perpendicular to the magnetization. 
Each of the two  targets has its magnetization aligned in the plane perpendicular to the x-ray propagation direction, at angles $\alpha_1=\pi/4$ and $\alpha_2=-\pi/4$ relative to the $\hat \pi$ direction. The more general case with arbitrary angles is discussed in Appendix~\ref{app:details}. In the linear regime, the nuclear scattering can then be described by  response functions $\Rlin^{(i)}(\omega), \Rcirc^{(i)}(\omega)$ for transitions with linear ($L$) and circular ($C$) dipole moments on target $i$~\cite{Siddons1999}. 
The $ \parapol$-polarized component of the outgoing electric field is then given by
\begin{align}
   \frac{ E_{\parapol}^\textrm{out}(\omega)}{E^{\textrm{in}}_{\perppol}(\omega)}  & = \frac{\mathcal{E}}{2}  \left [  \Rlin^{(1)}(\omega) \Rcirc^{(2)}(\omega) - \Rlin^{(2)}(\omega) \Rcirc^{(1)}(\omega) \right ] \,. \label{dark-fringe}
\end{align}
Here, $E^{\textrm{in}}_{\perppol}(\omega)$ is the input field, $\mathcal{E} = \exp(i k z - \mu_e\,d/2)$ and $\mu_e$ is the electronic contribution characterizing the absorption and phase shift due to off-resonant electronic processes. For identical targets with $\Rlin^{(1)} = \Rlin^{(2)}$ and $\Rcirc^{(1)} = \Rcirc^{(2)}$, the output signal $ E_{\parapol}^\textrm{out}(\omega)= 0$, such that the the dark-fringe condition is satisfied.

\subsection{Interfering polarization pathways}
Next, we explain the inner working principle of  the interferometer. It is instructive to focus on one of the hyperfine transitions.  As the six spectral lines are well separated, for a single transition with linear [circular] dipole moment at frequency $\omega_\mathrm{L}$ [$\omega_\mathrm{C}$], we can approximate the other response function as $R_\mathrm{C}(\omega_\mathrm{L})\approx 1$ [$R_\mathrm{L}(\omega_\mathrm{C})\approx 1$]. Then,
\begin{align}
        E_{\hat{\pi}}^\textrm{out}(\omega_\mathrm{L/C}) =& \frac{\mathcal{E}}{4}E_{\hat{\sigma}}^\mathrm{in}\left[\pm 2T_\mathrm{L/C}^{(2)} \mp 2 T_\mathrm{L/C}^{(1)} \right. \nonumber \\
     & \left. \pm T_\mathrm{L/C}^{(2)}T_\mathrm{L/C}^{(1)} \mp T_\mathrm{L/C}^{(2)}T_\mathrm{L/C}^{(1)} \right] \nonumber \\
     =& \pm \frac{\mathcal{E}}{4}E_{\hat{\sigma}}^\mathrm{in}\left[2T_\mathrm{L/C}^{(2)} - 2 T_\mathrm{L/C}^{(1)} \right]\,, \label{path}
\end{align}
where we have split the target response  $R^{(i)}_\mathrm{L/C} = 1 + T^{(i)}_\mathrm{L/C}$ into unscattered (1) and scattered ($T$) contributions, and 
left out the arguments $\omega_\mathrm{L/C}$ of $T$ for brevity. The different signs correspond to the $L/C$, depending on the chosen hyperfine transition. 

The four scattering contributions to the $\hat \pi$ polarization behind the second target can be read off from Eq.~(\ref{path}). These are indicated by the solid lines in Fig.~\ref{fig:setup}. The corresponding paths for the $\hat \sigma$ polarization are indicated as dashed lines in this figure. The overall nine scattering channels can be understood by noting that in each target, the electric field can pass without scattering, or can be scattered with (empty circle) or without (filled circle) change in linear polarization. 

Interestingly, we find that all multi-scattering pathways involving radiative coupling between the two targets cancel each other, even for two non-identical targets. The final interference signal leaving the interferometer only comprises two pathways, which correspond to (polarization-changing) scattering in either the first, or the second  target, respectively. Due to the absence of the usual radiative couplings, we can directly identify the two interfering paths with those in a conventional interferometer.

Note that since the two interfering pathways differ in polarization in between the two targets,  the dark-fringe interferometer  can be used to explore samples imposing polarization-dependent phase shifts, such as in circular dichroism~\cite{Vaz2025}. 

\subsection{Dynamical interferometer control}

We now show how the relative phase of the two interfering pathways can be dynamically controlled. For this, we assume that the first target is mounted on a piezo-electric transducer such that it can be mechanically displaced, thereby inducing relative phase shifts. In order to derive analytical results, we model the effect of the motion of the first target by instantaneous displacements $\Delta z$ immediately after the excitation. This approximation is valid as typical experimental rise times of the motion are of order of $1-10$~ns, which can be considered sufficiently short as compared to the nuclear lifetime of $141$~ns. 
For temporally short incident x-ray pulses $E^{\mathrm{in}}\propto\delta(t)$, such motion is characterized by the replacement 
$\Rlc(\omega)^{(1)} \to \Rlc^{(1)}(\omega, \phi) = 1 + \Tlc^{(1)}(\omega)\,e^{i \phi}$,
where $\Rlc^{(1)}(\omega, \phi)$ denotes the response with a motion-induced phase jump  $\phi = 2\pi \Delta z / \lambda$~\cite{Helistoe1991,Schindelmann2002,heeg2021coherent}. With this phase control, for identical targets, we obtain 
\begin{equation}
    E_{\hat{\pi}}^\textrm{out}(\omega_\mathrm{L/C}) = \frac{\mathcal{E}}{2}E_{\hat{\sigma}}^\mathrm{in} \left(1-e^{i\phi}\right)\, T_\mathrm{L/C}(\omega_\mathrm{L/C}) \, .
\end{equation}
Since the  spectral lines are well-separated, only one of the two $\Tcirc(\omega)$ or $\Tlin(\omega)$ contributes significantly to the detection signal at any given frequency, i.e., $\left|\Tcirc(\omega) - \Tlin(\omega)\right|^2 \approx |\Tcirc(\omega)|^2 +  \left|\Tlin(\omega)\right|^2$. Thus the above expression can be generalized to all hyperfine transitions, and the total signal intensity becomes
\begin{align}
 \frac{I_{\hat{\pi}}^\textrm{out}(\omega)}{I_{\hat{\sigma}}^{\textrm{in}}} 
 &=|\mathcal{E}|^2\,\sin^2(\phi/2)\:\left|\Tcirc(\omega) - \Tlin(\omega)\right|^2\,. \label{eq:int-out2}
\end{align}
As a result, the displacement phase $\phi$ allows one to continuously tune the intensity passing through the polarimeter, without affecting the spectral response of the targets. Note that the detection signal only comprises the pure nuclear response $\Tlc(\omega)$, since the off-resonant background light is blocked by the analyzer. Numerical simulations for different switching times, also including the resulting spectra of the outgoing x-rays, are discussed in the Appendix~\ref{app:numerics} and support these results.

\subsection{Experiment}
In order to explore the dynamically controllable interferometer and the x-ray intensity-gating, we have performed an experiment at the High Resolution Dynamics beamline P01 at the synchrotron PETRA III (DESY, Hamburg)~\cite{Wille_2010}. After passing the beamline's high-resolution monochromator, the x-ray pulses  were linearly polarized in the horizontal plane with the first crystal of the polarimetry setup at P01~\cite{Marx-Glowna2021}. The two targets are  $\sim 2\,\mu$m thick foils of $\alpha$-iron enriched in ${}^{57}$Fe to 95\%. The first foil is glued onto a  piezoelectric transducer in order to allow for a controlled motion. Two permanent magnets were used to align the respective internal magnetic fields of the samples. Subsequently, the x-rays were filtered by the polarimeter analyzer and detected by avalanche photo diodes. More details of the setup are provided in Appendix~\ref{app:setup}.

\begin{figure}[t]
 \centering
 \includegraphics[width=0.9\linewidth]{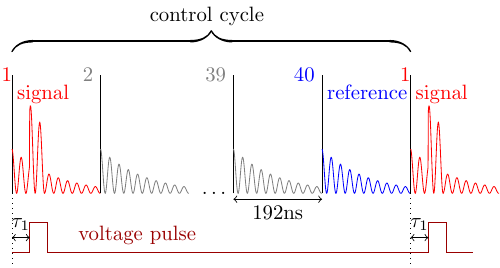}
 \caption{Schematic of the measurement approach in the experiment. We define a set of 40 subsequent x-ray pulses (indicated by black vertical lines, separated in time by 192ns) as one control cycle, corresponding to one revolution of the storage ring operated in 40-bunch mode. We record the sample response (depicted as exponentially damped and modulated signal) as function of time following the x-ray excitations.  In each control cycle, after the first x-ray pulses, we apply a voltage pattern to the piezo transducers inducing a motion. We denote this part of the control cycle  as the signal bunch $1$ (shown in red). The bunch before the signal bunch, i.e. the 40th bunch of the previous cycle, is used as a ``reference bunch'' in order to evaluate the signal in the absence of motion (shown in blue).
The applied voltage pattern comprises a single rectangular pulse (dark red, bottom part of figure). The start time of the voltage pattern relative to the arrival time of the synchrotron pulse $\tau_1$ can be varied.
For each voltage pattern, we repeat the control cycle many times and average the sample response in each of the 40 bunches of the control cycle separately. This way, we can analyze the effect of the target motion on different time scales, from nanoseconds within the signal bunch up to $40\times 192$\,ns across the entire control cycle.
 \label{fig:40bunchSequence}}
\end{figure}

\begin{figure}[t]
 \centering
    \includegraphics[width=0.9\linewidth]{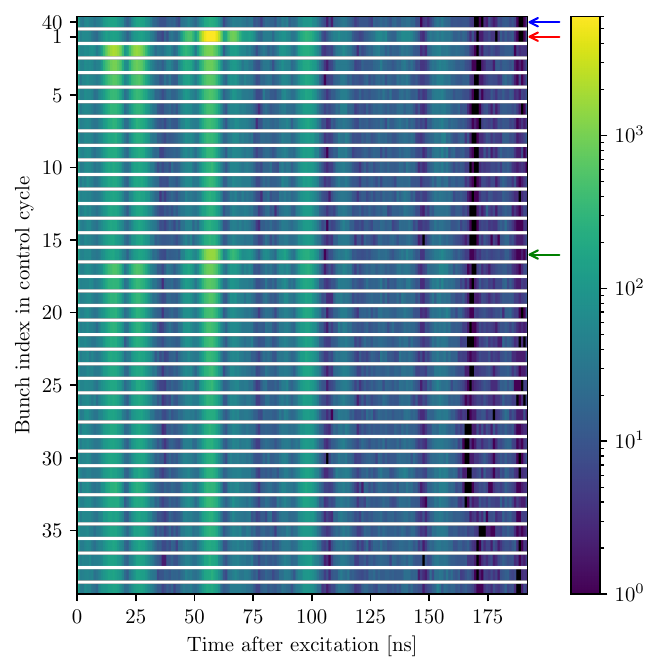}
    \caption{Experimentally recorded intensity of the nuclear forward scattering behind the analyzer as function of time, shown in color-coding on a logarithmic scale. The data acquisition followed    the NFS signals from the 40 bunches orbiting on the PETRA III circumference in the 40-bunch mode of operation, (corresponding to a duration of $7.68\,\mu$s), which was repeated many times. Each row in the figure shows the time-dependent intensity in the 192\,ns after one of the x-ray pulses in the cycle, such that the total cycle time of $7.68\,\mu$s runs from the top left to the bottom right in the figure. The target motion was applied following  the first x-ray pulse in the cycle, indicated by the red arrow. The x-ray signal from the bunch before, indicated by the blue arrow, is used as a reference. In the 16th pulse in the cycle, indicated by the green arrow, the intensity pattern shows a revival. The start time of the voltage pulse is 32\,ns after the arrival of the x-ray pulse. The corresponding data for the signal and reference bunch can also be found in Fig.~\ref{fig:time-spectra}.
    \label{fig:40traces}}
\end{figure} 

The storage ring was operated in 40-bunch mode, with $192$\,ns separation in between successive x-ray pulses. In order to study the effect of the sample motion in detail on different time scales, we induced sample motions by applying control-voltage patterns to the piezo transducer in every 40th x-ray pulse. During the remaining 39 excitation periods of 192\,ns each, no voltage was applied to the transducer. This pattern is shown schematically in Fig.~\ref{fig:40bunchSequence}, in which the ``signal bunch'' with sample motion is labeled by the index $1$. Within this control cycle, using an event-based detection system, we then measured the intensity of the scattered x-ray light following each of the 40 x-ray pulses separately. For the later analysis of the effect of the motion, we use the $40$th bunch, which is the bunch before the signal bunch, as the reference bunch, as it is expected to be least affected by the last piezo motion 39 bunches  earlier. This measurement approach allowed us to study not only the short time scales after a single x-ray excitation, but also the longer-term dynamics of the interferometer signal over 40 x-ray pulses ($7.68\mu$s).

In the main part of the experiment, we applied single rectangular voltage pulses of duration $20$\,ns at different start times $\tau_1$ after the arrival of the x-ray pulse, as schematically shown in the lower part of Fig.~\ref{fig:40bunchSequence}.  By studying the effect of the resulting motion of target 1 on the transmitted light, we can characterize the performance and the dynamical  control of the dark-fringe interferometer. 
Corresponding results  for two subsequent rectangular pulses with opposite polarity are discussed in Appendix~\ref{app:additional}.

In Fig.~\ref{fig:40traces}, we show the time-dependent intensities recorded for each of the 40 x-ray pulses in the control cycle separately, summed over many repetitions of the control cycle. The data is arranged such that the reference spectrum has bunch index 40 (indicated by the blue arrow), and is located directly above the signal spectrum with index 1 (indicated by the red arrow). There are three main effects visible in the data. First, the intensity in the signal spectrum 1 is modified by the motion, as compared to the control spectrum with index 40. This is a clear indication of the dynamical control of the interferometer. Second, revivals of these changes appear, best visible in the spectrum with index 16 (green arrow). These revivals will be traced back to mechanical waves in the support structure below the target. Finally, we observe background intensity in all time spectra,  which we attribute to residual relative motion between the samples as we will discuss below. 

\begin{figure}[t] 
\centering
    \includegraphics[width=\columnwidth, height=5cm]{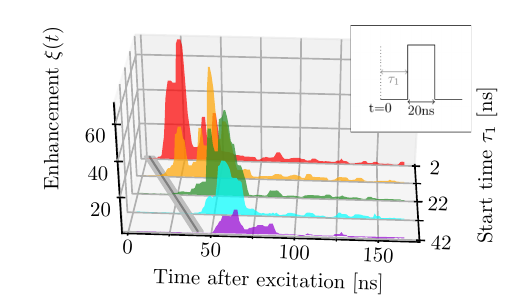}
    \caption{Experimental results for the x-ray intensity gating using the dynamically-controllable x-ray interferometer. The enhancement $\xi(t)$  defined in Eq.~(\ref{eq:enhancement}) of the main text is shown for different start times $\tau_1$ of the piezo motion after the x-ray excitation. The respective starting times are indicated by the dark gray diagonal line. It can be seen that the enhancement strongly increases after the start of the motion, demonstrating the possibility to switch on the intensity of the gated light via the motion-induced phase control.  After the applied voltage pattern has ended the enhancement quickly reduces back to low values, indicating the capability to gate the intensity by switching it on for a limited period only.
    \label{fig:gating_a}}
\end{figure} 

\begin{figure}[t] 
\centering
    \includegraphics[width=\linewidth]{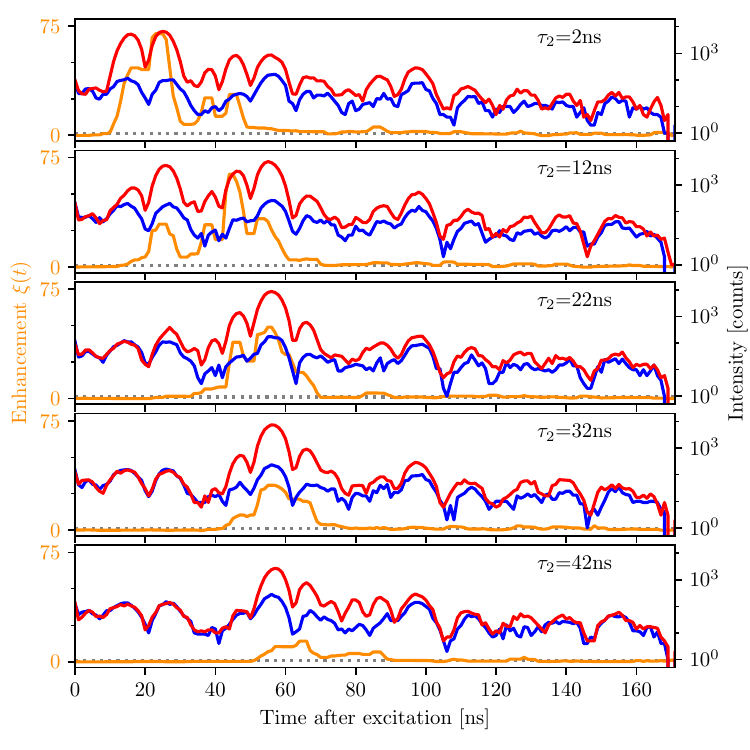}
    \caption{Time-dependent intensities recorded in the experiment. In each panel, the experimentally recorded raw signal spectrum is shown in red, and the corresponding reference spectrum in blue. In each sub-panel, the red and blue curves share the same respective intensity y-axis scale denoting the counts as in Fig~\ref{fig:40traces}. From each pair of experimental spectra, the enhancement is calculated as discussed in the main text, and shown in orange. The gray dashed line corresponds to an enhancement of 1. Note that the y-axes of the time spectra are scaled logarithmically, whereas the enhancement is shown on a linear scale. The measurements are performed with a single voltage pulse applied directly to the piezoelectric transducer as in Fig~\ref{fig:gating_a}.
    \label{fig:time-spectra}}
\end{figure} 

\subsection{Dynamical intensity gating}
We start by analyzing the dynamical intensity gating. In order to characterize the performance of the temporal gating, we define the enhancement
\begin{equation} \label{eq:enhancement}
  \text{enhancement}(t)=\frac{\text{signal}(t)}{\text{reference}(t)} =: \xi(t)
\end{equation}
given by the ratio of the time-dependent intensity in the signal bunch with motion to that in the reference bunch without motion. A high value in this ratio indicates the switching-on of the intensity in the signal bunch due to the interference control as compared to the reference bunch without motion. Results are shown in Fig.~\ref{fig:gating_a} for five different starting times $\tau_1$ of the voltage pattern after the exciting x-ray pulse. The respective starting times are indicated  by the diagonal dark gray line on the bottom of the plot. 
In all cases, the effect of the switching-on and switch-off of the intensity is clearly visible. Before the start of the voltage pattern, the enhancement is low. Thus, the polarimeter blocks the transmitted light up to experimental imperfections. After the start of the voltage pattern, immediately an increase of the enhancement is visible, indicating the desired controlled change of the phase-difference in the two paths to achieve  constructive interference. The individual time-dependent intensities entering the enhancement in Fig.~\ref{fig:gating_a} are displayed in Fig.~\ref{fig:time-spectra}.
Interestingly, the enhancement as defined in Eq.~(\ref{eq:enhancement}) exhibits a pronounced temporal dynamics approximately following the unperturbed time-dependent intensity  of the signal and reference. This can be understood by including a residual background intensity into the enhancement factor (see Appendix~\ref{app:residual}).

\begin{figure}[t] 
\centering
    \includegraphics[width=\columnwidth]{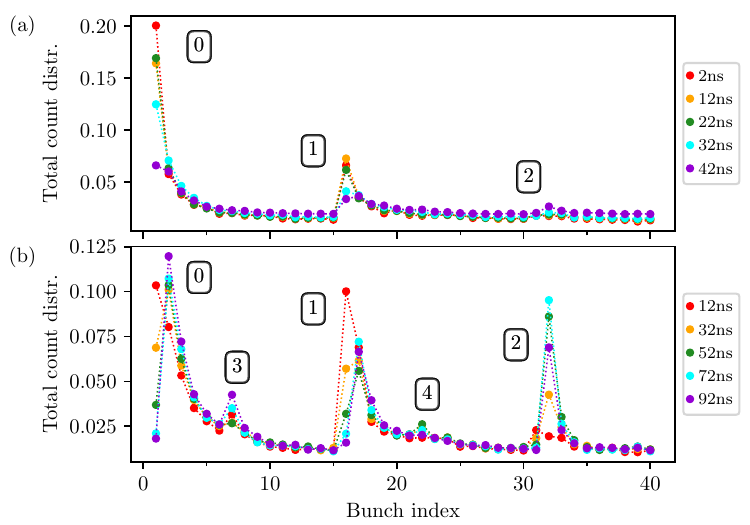}
    \caption{Time-integrated delayed intensity distribution as function of bunch index within the control cycle. The piezo movement is applied in bunch 1, leading to a rapid increase due to the intensity gating. Afterwards, periodic intensity revivals appear, which are labeled with increasing number in their order of appearance.  (a) shows the data corresponding to Fig.~\ref{fig:gating_a}, while (b) shows an analogous measurement with additional amplifier to increase the amplitude of the piezo motion. The different colors indicate the start time $\tau_1$ of the respective voltage  indicated on the right.
    \label{fig:integrated-counts}}
\end{figure} 

\subsection{Propagation of mechanical waves in the target}
Next, we explore the measurement capabilities of the interferometer. To this end, we analyze the "revivals" visible in Fig.~\ref{fig:40traces}, e.g., at bunch 16 indicated by the green arrow. To this end, we sum over the delayed photons recorded in each of the 40 bunches of the experimental sequence separately, and plot this intensity as function of bunch number. Results are shown in Fig.~\ref{fig:integrated-counts}(a). Upon applying the sudden displacement in bunch 1, the delayed intensity suddenly rises to the maximum value, which depends on the start time $\tau_1$. In the subsequent bunches, the intensity rapidly decays back to the background value. Interestingly, this decay is periodically interrupted by revivals of the integrated counts. They are labeled by their order of appearance using the rectangular labels in Fig.~\ref{fig:integrated-counts}. A closer analysis reveals that such revivals appear approximately every $15$ bunches. We attribute them to sample motion arising from a mechanical wave packet, which is created by the initial piezo kick and propagates through the acrylic glass plate support on which the piezo is mounted. Subsequently, it is reflected on the backside surface and returns to the sample. The resulting sample displacement induces the revival of the observed intensity. To establish this connection, we estimate the time between two revivals $j+1$ and $j$ ($j\in\{0, \dots, 4\}$) as
   $t_{j+1} - t_{j} \approx 15 \times 192\text{ns}  = 2.88 \mu s$,
 and compare it to the estimated roundtrip time of such a wave-packet
  $(2\times d)/v_s \approx 2.92 \mu s$,
where $d \approx 4\mathrm{mm}$ is the thickness of the acrylic glass plate, and $v_s \approx 2.74\times 10^3 \text{m}/\text{s}$ is the sound velocity in PMMA at room temperature~\cite{https://doi.org/10.1002/polb.22233}. The difference of approximately 40~ns between these two times is comparable to the uncertainty in $d$ and well below the integration time of 192~ns. 
To further verify the interpretation of a traveling mechanical wave, we repeated the experiment with a stronger piezo kick, using an amplifier to increase the applied voltage. The result is shown in Fig.~\ref{fig:integrated-counts}(b). In this case, up to four revivals of the stronger mechanical wave can be identified. This allows to study the sound wave propagation for more than 10\,$\mu$s, thus even exceeding the duration of the 40-bunch control cycle. Nevertheless, the revivals retain their period, in agreement with the above interpretation. This shows that the interferometer indeed is capable of detecting miniscule relative spatial motion between the targets on nanosecond time scales.

\begin{figure}[t]
 \centering
 \includegraphics[width=\columnwidth]{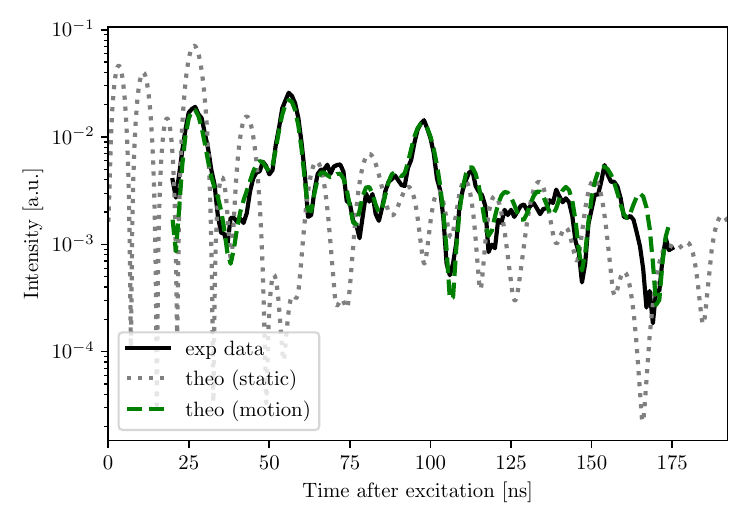}
 \caption{Background time-dependent intensity behind the analyzer, measured in the absence of voltages applied to the piezo transducer. In the ideal case of identical samples and without relative motion, this intensity is predicted to be zero. The non-zero measured intensity in particular contains information on the residual sample motion. The black curve shows the measured data. 
 The gray dotted curve shows corresponding theory data calculated in the static case, as explained in the main text. The green dashed curve shows corresponding theory data obtained assuming that the first target experiences random residual motions and that those residual motions are Gaussian distributed. The width of the Gaussian is fitted to be $\sigma_\textrm{det} = 0.302 \,\gamma$.}
 \label{fig:fitDetuning}
\end{figure}

\subsection{Residual sample motion}
Finally, we analyze the residual sample motion giving rise to the background signals visible in Fig.~\ref{fig:time-spectra}. For this,  we performed a reference measurement without applying a control voltage of the time-dependent intensities of each target separately, and of both targets in  the stationary  dark-fringe setting. We then compared the measured static interferometer intensity as function of time in Fig.~\ref{fig:fitDetuning} to model calculations. To this end, we employ the software package {\sc Pynuss}~\cite{pynuss} to fit material models to the time-resolved spectra of the two targets individually. From these models, the combined response of the two targets is calculated in the absence of motion.  It can be seen that the experimental data (solid black line) cannot be described adequately using this theoretical model approach without any motion of the targets (gray dotted line). 

These deviations between the theory model and the experimental data can be attributed to residual relative motions of the two targets. In order to show this, we assume that the relative velocity of the two targets changes slowly, such that it can be approximated as being constant throughout  the 192\,ns measurement time scale following each x-ray excitation. This allows us to  model the relative velocity via a Doppler-detuning $\Delta_\textrm{rel}$ of one of the targets with a Gaussian distribution centered around 0. A fit of the width of this distribution yielded a best result with  a standard deviation of $\sigma_\textrm{det}=0.302\,\gamma$, where $\gamma=4.7\,$neV is the linewidth. The theoretically predicted intensity taking into account  this residual sample motion is shown as the green dashed line in Fig.~\ref{fig:time-spectra} and agrees well with the measured data. More details on this analysis are given in Appendix~\ref{app:residual}.

\section{Discussion}

In summary, we introduced and demonstrated an inline x-ray interferometer for M\"ossbauer science operated in the dark-fringe mode.
The resulting minimum of the transmitted intensity  forms an ideal starting point for a variety of applications. Mounting one of the two targets on a piezo-transducer further allows one to dynamically control the relative phase of the interfering pathways, and thus the interferometer transmission.

The dark-fringe operation is realized by operating two targets containing M\"ossbauer nuclei in a particular geometry, such that their individual scattering responses partially interfere. Importantly, all scattering channels involving interactions with both targets, in  contrast to previous setups involving multiple M\"ossbauer targets. Due to this feature, the final signal behind the analyzer only comprises two interfering contributions, which are formed by individual scatterings from each of the two targets. The two contributions have orthogonal polarizations between the two targets, such that they can be used for probing other samples. We then showed that the relative phase between the two pathways can dynamically be controlled via the mechanical displacement of the corresponding sample.  These  features allow us to directly relate the setup to a conventional interferometer with two interfering pathways.

In the experiment, we then demonstrated the capabilities of the interferometer  by dynamically controlling the  intensity transmitted through the interferometer, and by precisely measuring relative target displacements. A notable feature of the experiment was the measurement over a 40-bunch sequence, which allowed us to study mechanical dynamics up to 7.68\,$\mu$s after the initial excitation. Only in this way, we could reveal the presence of mechanical sound waves excited by the sudden piezo motion and traveling through the target for extended times.

For the future, we envision a more refined control of the mechanical displacement to generate versatile tuneable multi-peak pulses. This would allow, e.g., for the coherent control of M\"ossbauer nuclei~\cite{heeg2021coherent}. If combined with seeded XFEL radiation providing many resonant photons per shot, Ramsey-like control operations or even multi-dimensional spectroscopy could come within reach. Such control could be achieved along the lines of adaptive optics, using feedback control to determine the piezo motion which  achieves the desired time-dependent x-ray pulses.  On the measurement side, the polarization-dependent nature of the pathways in the dark-fringe interferometer could be harnessed to study polarization-dependent effects, such as circular dichroism, in a space- and time-resolved way.
An interferometry of the type reported here is also a key requirement for protocols to  verify x-ray single-photon entanglement~\cite{PhysRevLett.103.017401}.
Furthermore, our approach to explore relative target motion with  nanosecond time-resolution and x-ray-wavelength spatial-resolution over extended measurement times of more than 10\,$\mu $s appears particularly suitable for the study of impulsively induced mechanical motion, e.g., initiated  by laser pulses or  x-ray scattering.

\begin{acknowledgments}
We acknowledge DESY (Hamburg, Germany), a member of the Helmholtz Association HGF, for the provision of experimental facilities. Parts of this research were carried out at PETRA III. Data was collected using the P01 High Resolution Dynamics Beamline operated/provided by DESY Photon Science. Beamtime was allocated for proposal I-20200886.
We gratefully acknowledge support in the design and installation of one of the magnets by Jos\'e Crespo and Conrad Hagemeister.
The polarimetry setup was funded by the BMBF grant no. 05K13SJ1.
\end{acknowledgments}

\appendix

\section{\label{app:setup}Experimental setup.}
The target motion was controlled by applying voltage patterns to a piezoelectric transducer on which the sample was glued. The applied voltage patterns were generated via an arbitrary-waveform signal generator (Keysight 81160A-002) and either applied to the piezo transducer directly or via an amplifier (ENI 3100 LA) / attenuator combination (+25dB in total) in order to generate larger piezo motion amplitudes. A 50\,$\Omega$ load resistor (Spinner BN 53 12 12) is mounted in parallel close to the piezo transducer in order to terminate the transmission line. The employed voltage patterns are schematically shown in the lower part of Fig.~5 in the main text and comprise either a single rectangular pulse or two subsequent rectangular pulses with opposite polarity. Each rectangular pulse has a duration of $20$\,ns, and the pulse amplitudes at the signal generator are 1.2V${}_{\mathrm{pp}}$ with and 5V${}_{\mathrm{pp}}$ without the amplifier. It is important to note, however, that the actual motion of the piezo transducer cannot be expected to follow the applied voltage pattern directly, due to the frequency-dependent response function of the transmission line comprising cables, the amplifier, the piezo itself, and the termination resistor, as well as the mechanical inertia of the moving sample. Throughout the experiment, we scanned the starting time $\tau_1$ of the voltage pattern relative to the x-ray pulse arrival. The voltage pulse was applied once every 40 bunches as  shown in Fig.~\ref{fig:40bunchSequence}. 
 
The two M\"ossbauer samples were fabricated from adjacent pieces of a $\sim 2\mu$m thick foil made from $\alpha$-iron enriched in ${}^{57}$Fe to 95\%. The first moveable foil is glued onto a thin polyvinylidene fluoride piezoelectric transducer (DT1-028K, Measurement Specialities Inc.) using cyanoacrylate instant adhesive.  The piezoelectric transducer in turn is glued to a $4$\,mm thick acrylic glass plate using two-part epoxy glue. The stationary foil is mounted between two Kapton foils. Two permanent magnets are used to align the respective internal magnetic fields of the samples. The magnet of the static sample is rotatable around the beam propagation axis using a motorized stage in order to adjust the relative orientation of the two magnetization directions. For the theoretical analysis, we characterized the samples via model fits to time-dependent intensities measured for individual targets. For the static target, the best fit was achieved for a thickness of $d=1.92\,\mu$m, and a magnetic field of $B=32.69$\,T with polar angle $\theta = 0.498\pi$ and azimuthal angle $\alpha = 0.255\pi$. For the moving sample, the corresponding fit parameters are $d=1.89~\mu$m, $B=32.65$\,T, $\theta = 0.483\pi$ and $\alpha = 0.750\pi$.

\section{\label{app:residual}Analysis of the residual sample motion.}

In Fig.~\ref{fig:fitDetuning} and the subsequent discussion, we have analyzed the residual sample motion in the absence of the voltage pulse, which leads to significant changes of the time spectrum. Assuming a slow noise motion allows us to  model the relative velocity via a Doppler-detuning $\Delta_\textrm{rel}$ of one of the targets, followed by an average of the theory predictions over a Gaussian distribution of detunings with width $\sigma_\textrm{det}$,
\begin{align}
p\left(\Delta_\textrm{rel}\right) =  \frac{1}{2\sqrt{\pi}\sigma_{\mathrm{det}}}\: e^{-\left(\Delta_\textrm{rel}/2\sigma_\textrm{det}\right)^2}\,.
\end{align}
Note that for ${}^{57}$Fe, a detuning of one natural linewidth approximately corresponds to a velocity of 0.1 mm/s.
The resulting averaged time-dependent intensity 
\begin{align} \label{eq:fitModel_onlyDet}
    I_\mathrm{avg}(t) = & \int_{-\infty} ^\infty \mathrm{d}\Delta_\textrm{rel} \: p\left(\Delta_\textrm{rel}\right)
    \left| \left[ e^{i \Delta_\textrm{rel}t }E_1(t) \right]\ast E_2(t)\right|^2 
\end{align}  
is then fitted to the experimental data, with the distribution width $\sigma_\textrm{det}$ as the fit parameter. Here, $E_1(t)$ and $E_2(t)$ are the amplitudes of the x-rays scattered by the two samples into the perpendicular polarization channel, respectively.

The result of this residual motion fit is shown as the green dashed curve in Fig.~\ref{fig:fitDetuning}, with $\sigma_\textrm{det} = 0.302 \,\gamma$. It can be seen that the theoretical result with residual sample motion agrees well with the experimental data over the entire time range. Therefore, we conclude that even in the absence of piezo motion induced by the control voltage, the distance between the two samples fluctuates throughout the experiment. 
 
The fitted detuning spread of $\sigma_\textrm{det} = 0.302 \,\gamma$ corresponds to a spread of velocities of $\sigma_v = c \sigma_\textrm{det} / E_\gamma$=0.03\,mm/s, where we used the speed of light $c$ and the photon energy $E_\gamma=14.4$\,keV.
For Gaussian vibration noise the spread of the velocity $\sigma_v$ and the spread of the displacement $\sigma_x$ are related by the frequency content. Assuming a sinusodial vibration $x(t)=A\cos\left(2\pi f t +\phi\right)$ with a characteristic frequency $f$, the spread of velocities $\sigma_v$ and the spread of displacements $\sigma_x$ are related by $\sigma_v\approx 2\pi f \sigma_x$. This formula can be easily derived from the variance of $x(t)$ and $v(t)$ and averaging over the phase $\phi$. For a typical vibration frequency for acoustic vibration of $f\approx 100-1000$\,Hz, we find a spread of displacements of $ \sigma_x \approx 50-500\,\text{nm}$. 
The susceptibility of the system to this frequency range was experimentally confirmed using test measurements with additional loud speakers playing music on ambient sound level directed onto one of the targets. In these tests,  a strong increase of transmitted intensity was observed, which can be attributed to an additional residual target motion induced by the sound waves. This again highlights the sensitivity of our setup to target motion due to the dark-fringe operation.
 
\section{\label{app:enhancement}Time-dependence of enhancement.}
In Fig.~\ref{fig:gating_a} we found that the  enhancement in the transmitted intensity induced by the dynamical interferometer-control is time-dependent. This can be  understood by analyzing the enhancement in the presence of background intensity contributions $\rho(t)$,
\begin{align}
\xi_{\mathrm{exp}}(t) = \frac{\text{signal}(t)+\rho(t)}{\text{reference}(t)+\rho(t)}\,.
\end{align}
At times of low signal and reference intensities,  the background $\rho(t)$ dominates and the enhancement on average approaches 1. Conversely, when the signal dominates, $\xi_{\mathrm{exp}}(t)\approx \xi(t)$, and the unperturbed enhancement is observed. As a result, the observed enhancement is expected to vary with the signal intensity, and therefore follows the time-dependent intensity of the scattered light.


\section{\label{app:details}Details on the dark-fringe interferometer}
\subsection{Polarization-dependent response of a moving target}
 We start with the analysis for a single target, assuming an idealized motion, which will allow us to obtain analytical expressions. For definitiveness, we consider the standard isotope ${}^{57}$Fe with resonance frequency $\omega_0$ and a magnetic dipole (M1) M\"ossbauer transition from the ground state.  Following the standard approach to polarization-dependent nuclear forward scattering~\cite{Siddons1999,ralf}, we can relate the linear $ \parapol$- and the $ \perppol$-polarized components of the outgoing electric field ($E$) to the incident electric field ($E^\textrm{in}$) by 
\begin{align}
  \begin{pmatrix}E_{\perppol}(\omega) \\ E_{\parapol}(\omega) \end{pmatrix} & = 
 \mathcal{E} \: \mathcal{R}(\omega, \alpha)
  \begin{pmatrix}E^{\textrm{in}}_{\perppol} \\ E^{\textrm{in}}_{\parapol} \end{pmatrix}\,,\label{eq:likejones}\\[2ex]
  \mathcal{R}(\omega, \alpha) &= g_\alpha  \begin{pmatrix} \Rlin(\omega) & 0\\0 & \Rcirc(\omega) \end{pmatrix} g^{-1}_\alpha\,, \label{r-static}
\end{align}
where $\mathcal{E} = \exp(i k z - \mu_e\,d/2)$ and $\mu_e$ is the electronic contribution characterizing the absorption and phase shift due to off-resonant electronic processes. $g_\alpha$ is a two-dimensional rotation matrix,
\begin{align}
 g_\alpha = \begin{pmatrix}
                \cos \alpha  & \sin \alpha  \\
                -\sin \alpha & \cos \alpha
            \end{pmatrix}\,,
\end{align}
capturing the effect of the magnetic field being rotated by $\alpha$ from the $\hat{\pi}$ axis.
For ${}^{57}$Fe, the two response functions for transitions with linear and circular dipole moments are given by
\begin{subequations}
\begin{align}
\Rlin(\omega) &= e^{\mathcal{L}_2(\omega)+\mathcal{L}_5(\omega)}\,,\\
\Rcirc(\omega)  &= e^{\frac34\mathcal{L}_1(\omega)+\frac14\mathcal{L}_3(\omega) +\frac14\mathcal{L}_4(\omega)+\frac34\mathcal{L}_6(\omega)}\,.
\end{align}
\end{subequations}
Here, the Lorentzians $\mathcal{L}_i(\omega) = i\Gamma_c/(\omega - \omega_i - \frac{i}{2}\gamma )$
correspond to the six Zeeman transitions in magnetically-split ${}^{57}$Fe, weighted by their respective Clebsch-Gordan coefficients, with transition frequencies $\omega_i$, natural decay rate $\gamma$ and enhanced decay rate $\Gamma_c$~\cite{Siddons1999}.

For simplicity, we consider instantaneous phase jumps due to step-like target displacements $\Delta z$ immediately after the excitation. This approximation is valid as typical rise times of the motion are of order of $1-10$~ns, which can be considered sufficiently short as compared to the nuclear lifetime of $141$~ns. For temporally short incident x-ray pulses $E^{\mathrm{in}}\propto\delta(t)$, the effect of these step-like motions is characterized by the replacement \cite{Helistoe1991}
\begin{align}\label{eq:response-with-motion}
\Rlc(\omega) \to \Rlc(\omega, \phi) = 1 + [\Rlc(\omega) - 1]\,e^{i \phi}\,,
\end{align}
where $\Rlc(\omega, \phi)$ denotes the response with phase jump  $\phi = 2\pi \Delta z / \lambda$. Note that the first ``1'' in Eq.~\eqref{eq:response-with-motion} describes unscattered light which does not interact with the target and therefore is  unaffected by the motion, whereas the resonantly scattered part proportional to $(\Rlc-1)=:T_\mathrm{L/C}$ is phase-shifted.
Note that as the six hyperfine-splitted lines are well-separated, specifying the transition to linearly polarized $\omega_\mathrm{C}$ or linearly polarized $\omega_\mathrm{L}$, we can approximate $R_{\mathrm{L/C}}(\omega_\mathrm{C/L})\approx 1$.

\subsection{Dynamical interference control}

In order to obtain a detection signature which allows one to directly observe the intensity gating in an experiment, we combine one moving target (magnetization $\alpha_1=\pi/4$) with an (ideally) identical {\it static} one ($\alpha_2=-\pi/4$). 
From Eq.~(\ref{eq:likejones}), for magnetization directions $\pm \pi/4$ the outgoing field comprises contributions in both polarization directions, even if the incident field is only comprises one of the two polarization components. This is due to polarization-changing scattering contributions~\cite{Siddons1999}.
In contrast, we combine two targets in such a way that this is not the case. After the analyzer, only the $\hat \sigma$ component of the short incoming x-ray pulse impinges on target 1. The electric field after the first target then becomes
\begin{align}
    E^{(1)}(\omega_\mathrm{L/C}) =& \mathcal{R}^{(1)}\left(\omega_\mathrm{L/C}, \phi, \alpha_1=\frac{\pi}{4}\right)
     \begin{pmatrix}E^{\textrm{in}}_{\perppol} \\ 0\end{pmatrix} \\
     =&\frac{\mathcal{E}}{2}E_{\hat{\sigma}}^\mathrm{in}
     \begin{pmatrix}
         2 + T_\mathrm{L/C}^{(1)}e^{i\phi} \\
         \mp T_\mathrm{L/C}^{(1)}e^{i\phi}
     \end{pmatrix}
\end{align}
where we left out the argument $\omega_\mathrm{L/C}$ of the transmission functions for brevity. The signs $\mp$ are for the two different polarization contributions $L/C$. From the equation, we can read off the three paths after the first sample as they are illustrated in Fig.~\ref{fig:setup}(b) in the main text: the prompt unscattered contribution ("1") and the scattering components into the parallel and perpendicular polarization components ("$T_\mathrm{L/C}^{(1)}e^{i\phi}/2$").

\begin{figure}[t]
\centering
    \includegraphics[width=0.95\columnwidth]{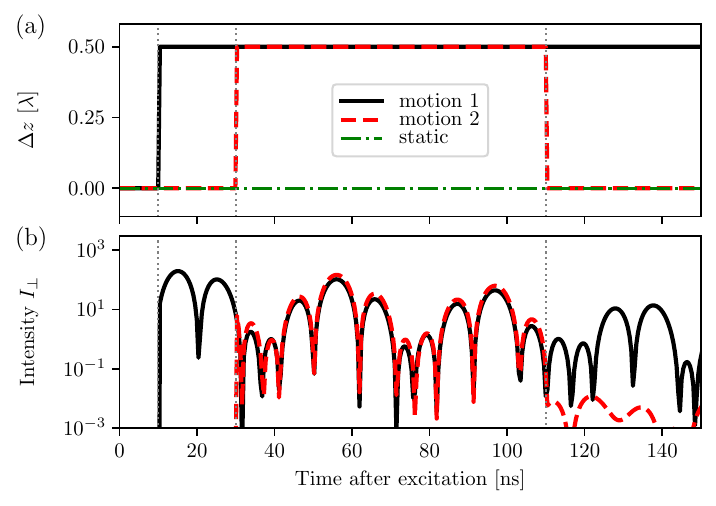}
    \caption{Dynamical x-ray intensity switching using mechanically induced linear polarization rotation together with a polarimetry setup (see Fig.~\ref{fig:setup} in the main text for a schematic of the setup). (a) shows three motions of target 1 considered in the analysis. Target 2 remains static in all cases. (b) shows the corresponding intensities registered behind the analyzer. Without motion, and before the first displacement (at $t=10$~ns and $t=30$~ns for motion 1/2, respectively), the intensity is zero. For motion 2, moving back at $t=110$ns again suppresses the outgoing intensity.   Both targets are $\alpha$-Fe foils of 1~$\mu$m thickness, with magnetization aligned along $\alpha_1 = \pi/4$ and $\alpha_2 = -\pi/4$, respectively. \label{fig:polarimeter}}
\end{figure}
\begin{figure}[t] 
\centering
    \includegraphics[width=0.9\columnwidth]{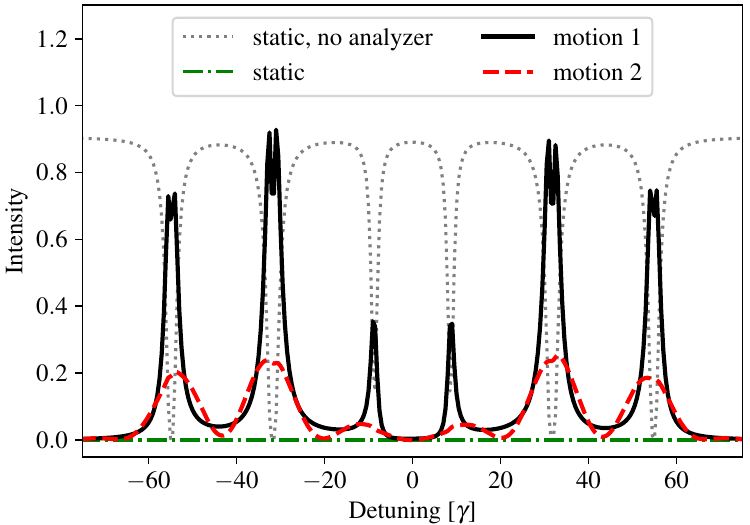}
    \caption{Spectrum of the dynamically switched x-ray light, where the detuning is that of the incident x-ray light frequency relative to the unsplit nuclear resonance frequency, in units of the single-nucleus linewidth $\gamma$. The spectra correspond to the time-dependent intensities shown in Fig.~\ref{fig:polarimeter}. Further, a spectrum without analyzer is shown as a reference, which shows the usual absorption dips at the positions of the six Zeeman transition frequencies in the spectrally broad background created by the prompt pulse.
    \label{fig:spectrum} 
    }
\end{figure} 
Analogously, after the second target we find
\begin{align}
 E^\textrm{comb}(\omega_\mathrm{L/C}) =& 
\mathcal{E} \: \mathcal{R}^{(2)}\left(\omega_\mathrm{L/C}, - \frac{\pi}{4}\right)\: \times \nonumber \\[2ex]
&\quad \times \mathcal{R}^{(1)}\left(\omega_\mathrm{L/C}, \phi, +\frac{\pi}{4}\right)
  \begin{pmatrix}E^{\textrm{in}}_{\perppol} \\ 0\end{pmatrix} \nonumber\\[2ex]
=& \frac{\mathcal{E}}{4}E_{\hat{\sigma}}^\mathrm{in}
\begin{pmatrix}
 S^{(0)} + S^{(1)} + S^{(2)}\\ 
P^{(1)} + P^{(2)}
\end{pmatrix}\,.
\end{align}
The different scattering contributions  are given by 
\begin{align}
S^{(0)} =& 4\,,\\[2ex]
S^{(1)} =& 2 \left( T_\mathrm{L/C}^{(1)}e^{i\phi} +  T_\mathrm{L/C}^{(2)} \right)\,,\\[2ex]
S^{(2)} =& T_\mathrm{L/C}^{(2)}T_\mathrm{L/C}^{(1)} e^{i\phi}-T_\mathrm{L/C}^{(2)}T_\mathrm{L/C}^{(1)}e^{i\phi}= 0\,,\\[2ex]
P^{(1)} =& \pm 2 \left( T_\mathrm{L/C}^{(2)} -  T_\mathrm{L/C}^{(1)}e^{i\phi} \right)\,,\\[2ex]
P^{(2)} =& \pm T_\mathrm{L/C}^{(2)}T_\mathrm{L/C}^{(1)}e^{i\phi}\mp T_\mathrm{L/C}^{(2)}T_\mathrm{L/C}^{(1)}e^{i\phi} = 0\,,
\end{align}
and each of these nine terms corresponds to one of the interfering pathways shown in Fig.~\ref{fig:setup}(b) of the main text. Here, $S/P$ indicates contributions to the $\hat \sigma$ and $\hat \pi$ polarization, respectively, and the superscript labels the number targets that this pathway interacts with.
%
Note that in both polarization components, for this particular set of chosen angles $\alpha_{1/2}$, the scattering paths involving two targets cancel each other, $S^{(2)} = 0 = P^{(2)}$ even for non-equal targets. 
The analyzer blocks the $\hat{\sigma}$-component such that the outgoing signal behind the interferometer becomes
\begin{align}
    E^\textrm{out}(\omega_\mathrm{L/C}) = \pm \frac{\mathcal{E}}{2}E_{\hat{\sigma}}^\mathrm{in}\left( T_\mathrm{L/C}^{(2)} -  T_\mathrm{L/C}^{(1)}e^{i\phi} \right)\, .
\end{align}
Assuming identical samples $T_\mathrm{L/C}^{(1)}=T_\mathrm{L/C}^{(2)}=T_\mathrm{L/C}$, the outgoing field reduces to
\begin{equation}
    E^\textrm{out}(\omega_\mathrm{L/C}) = \pm \frac{\mathcal{E}}{2}E_{\hat{\sigma}}^\mathrm{in}(1-e^{i\phi})T_\mathrm{L/C}(\omega_\mathrm{L/C}) \, .
\end{equation}
Without motion ($\phi=0$), this outgoing field indeed vanishes in the ideal case due to destructive interference between the respective single-scattering paths, thus forming the dark-fringe setting. As a result, the interference and thus the outgoing intensity 
\begin{align}
 \frac{I_\perp(\omega_\mathrm{L/C})}{I_{\textrm{in}}} 
 &=|\mathcal{E}|^2\,\sin^2\left(\frac{\phi}{2}\right)\:\left|T_\mathrm{L/C}(\omega)\right|^2\,. \label{eq:int-out}
\end{align}
can be dynamically controlled, e.g., by  mechanical motion inducing a relative  phase $\phi$.

\section{\label{app:numerics}Numerical analysis of the interferometer performance}

\begin{figure}[t!] 
\centering
    \includegraphics[width=\linewidth]{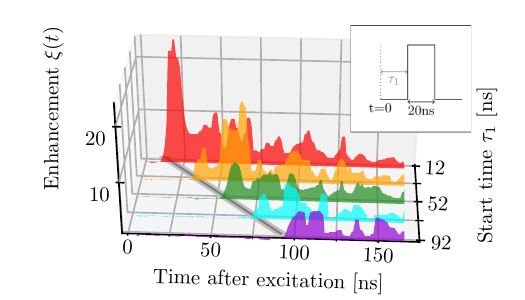}
    \caption{Experimental results for the motion-induced enhancement with single voltage pulse shown in the inset. The respective traces show the enhancements for different delays $\tau_1$ between the voltage pulses and the incident x-ray pulse, with the starting time of the voltage pulse indicated by the thick gray line.  The results are recorded with amplifier.
    \label{fig:single-pulse}}
\end{figure} 
Exemplary results of a full numerical simulation of the setup   are shown in Fig.~\ref{fig:polarimeter}. We compare the three motions shown in panel (a): the static case, one displacement at $t=10$~ns, and two consecutive displacements of opposite magnitude at $t=30$~ns and $t=110$~ns.
Panel (b) shows the corresponding x-ray intensity behind the analyzer as a function of time. The temporal gating of the signal due to the dynamical polarization rotation is clearly visible. In case of motion 1, the first 10~ns of the signal before the displacement are suppressed. For motion 2, x-rays outside the interval 30~ns--110~ns are suppressed. The intensity in case of the static sample is zero and therefore not visible in the logarithmic plot.  

Fig.~\ref{fig:spectrum} shows the corresponding frequency spectra of the light passing the polarimeter. These indeed confirm that the transmitted light resembles the usual six-line spectrum of ${}^{57}$Fe, but without the off-resonant background.  Interestingly, the resonant intensities may approach or even exceed the incoming light intensity, depending on the duration of the interval in which the x-rays may pass the setup. Furthermore, the comparison of the spectra of motions 1 and 2 in Fig.~\ref{fig:spectrum} shows that  shorter pass durations lead to a broadening of the spectral lines, consistent with the usual Fourier relations. 

\begin{figure}[t] 
\centering
    \includegraphics[width=\linewidth]{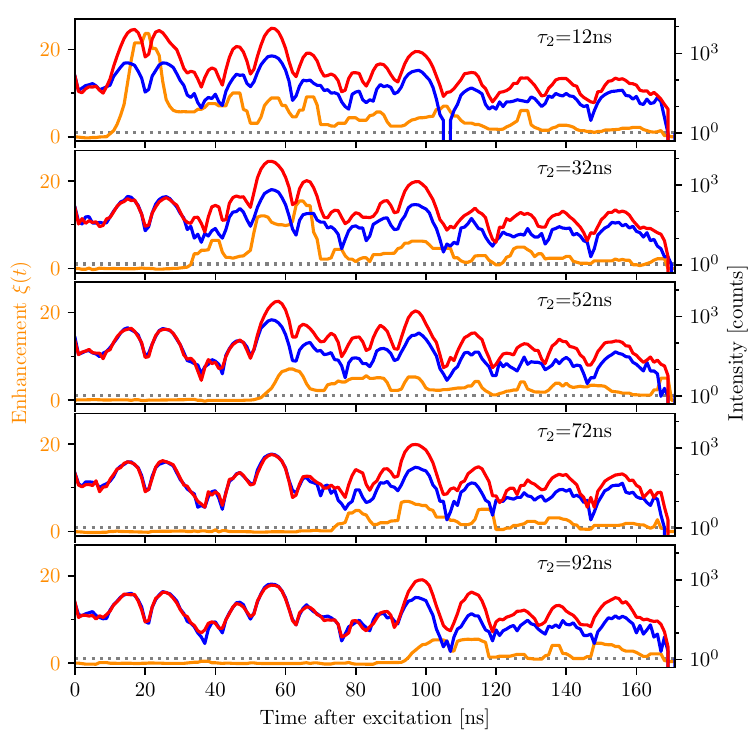}
    \caption{Time-dependent intensities recorded in the experiment with the single pulse voltage sequence and amplifier. In each panel, the experimentally recorded raw signal spectrum is shown in red, and the corresponding reference spectrum in blue. In each sub-panel, the red and blue curves share the same respective intensity y-axis scale denoting the counts of the bunch with motion (red) and the reference bunch (blue). From each pair of experimental spectra, the enhancement is calculated as discussed in the main text, and shown in orange. The gray dashed line corresponds to an enhancement of 1. Note that the y-axes of the time spectra are scaled logarithmically, whereas the enhancement is shown on a linear scale. In the different rows the delay time of the voltage pulse $\tau_1$ is varied.
    \label{fig:time-spectra_single}}
\end{figure}

\section{\label{app:additional}Additional measurements}

\begin{figure}[t] 
\centering
    \includegraphics[width=\linewidth]{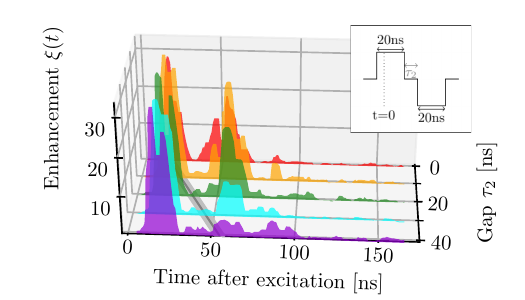}
    \caption{Experimental results for the motion-induced enhancement with double voltage pulses shown in the inset. The respective traces show the enhancements for different delays $\tau_2$ between the two voltage pulses of opposite polarity, with the starting time of the second voltage pulse indicated by the thick gray line. The first voltage pulse start time is $\tau_1=-5\,$ns, such that the motion is initiated shortly before the synchrotron pulse arrival at time $t=0$. The results are recorded without amplifier.
    \label{fig:double-pulse}}
\end{figure} 

\begin{figure}[t] 
\centering
    \includegraphics[width=\linewidth]{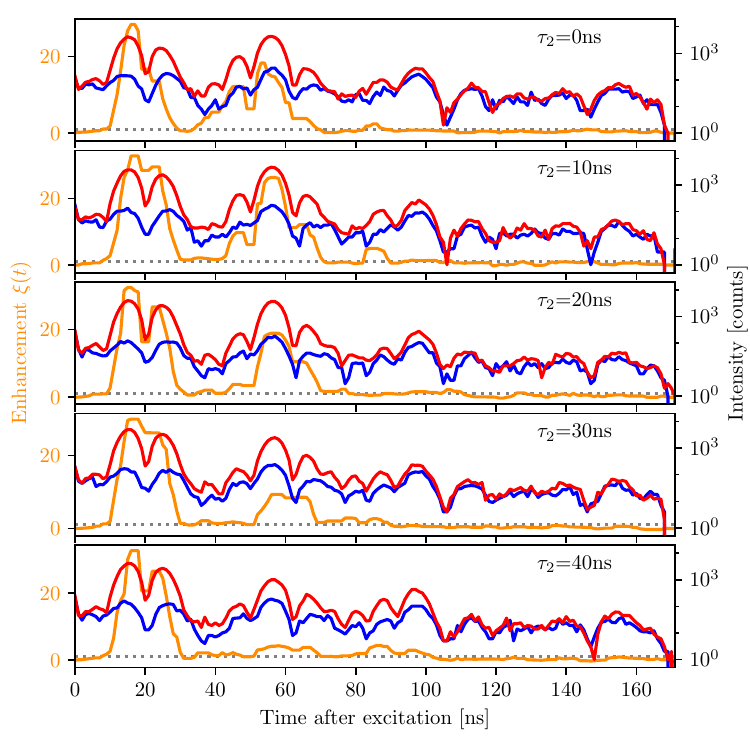}
    \caption{Time-dependent intensities recorded in the experiment with the double pulse voltage sequence. In each panel, the experimentally recorded raw signal spectrum is shown in red, and the corresponding reference spectrum in blue. In each sub-panel, the red and blue curves share the same respective intensity y-axis scale denoting the counts of the bunch with motion (red) and the reference bunch (blue). From each pair of experimental spectra, the enhancement is calculated as discussed in the main text, and shown in orange. The gray dashed line corresponds to an enhancement of 1. Note that the y-axes of the time spectra are scaled logarithmically, whereas the enhancement is shown on a linear scale. In the different rows the time between the pulses $\tau_2$ is varied.
    \label{fig:time-spectra_double}}
\end{figure} 

\subsection{\label{sec:singleWith}Single voltage pulses with amplifier}
In the main text, the pulse gating results were discussed for measurements with a single voltage pulse applied directly to the piezo electric transducer. In the experiment, also further measurements were performed with an additional amplifier to increase the piezo motion (see Appendix~\ref{app:setup}). The corresponding enhancements and time-dependent intensities are shown in Figs.~\ref{fig:single-pulse} and \ref{fig:time-spectra_single}. In all cases, the onset of the enhancement is clearly visible. However, compared to the measurements without amplifier, the enhancement is smaller. We attribute this to a higher background motion level, leading to a higher intensity already in the reference bunch. This is consistent with the discussion of Fig.~\ref{fig:integrated-counts} in the main text, where it is shown that with amplifier, the induced sound waves decay slower such that  this residual motion gives a higher background. Furthermore, although the voltage pulse generated by the signal generator has the same shape, in the measurements with amplifier we see enhancement for a significantly longer time. This is most probably caused by frequency-dependent amplification, resulting  in a different pulse shape  after the amplifier. We therefore conclude that even though the time-dependent intensities for the signal bunch, in which the voltage pulse is applied, have a similar intensity level, the cleaner background and better pulse control make the measurements without amplifier more favorable for a dynamical pulse-gating.

\subsection{\label{sec:double}Double voltage  pulses}
In the main text, we focused on voltage patterns comprising a single rectangular pulse. Here, we present additional results obtained with two voltage pulses of opposite polarity. Results are shown in Fig.~\ref{fig:double-pulse}, where  $\tau_2$ is a variable delay between the two pulses. In all cases, the starting time of the voltage pulse sequence is fixed to $\tau_1=-5\,$ns, i.e.~it starts 5~ns before the x-ray pulse arrives. This value was chosen in order to compensate for the finite time of the initial motion after application of the voltage. The first voltage pulse leads to similar enhancements for all considered settings of $\tau_2$, analogously to the single-pulse cases in Fig.~\ref{fig:gating_a} in the main text. This correspondence is expected, as the first pulse does not depend on the value of $\tau_2$. In contrast, the motion-induced enhancement after this first pulse varies with $\tau_2$ in a systematic way. In particular, a region of lower enhancement appears after the first enhancement pulse, with duration increasing with $\tau_2$. However, it is not possible to clearly identify two separated pulses with constant shape and low or even zero enhancement with duration depending on $\tau_2$ in between. In particular, the shape of the second pulse part varies with $\tau_2$. We attribute this mostly to residual sample motions following the first voltage pulse, throughout the delay time $\tau_2$, which impede a well-defined gating of the second pulse. As discussed and observed for the single-pulse voltage patterns, the motion-induced enhancement also follows the time-dependent intensity of the reference intensity, which is shown in Fig.~\ref{fig:time-spectra_double}, itself. This dependence also renders a clear observation of well-defined double-pulses more challenging. This motivates the development of better-defined motional control in order to open up the possibility to dynamically structure x-ray pulses, e.g., with tuneable delays.


\bibliography{literatur}

\end{document}